\documentclass[]{spie}  

\usepackage{amsmath,amsfonts,amssymb}
\usepackage{graphicx}
\usepackage{xcolor}
\usepackage[colorlinks=true, allcolors=blue]{hyperref}
\usepackage[caption=false]{subfig}
\usepackage{nicefrac, pbox, ltablex}
\usepackage{lineno}

\begin{document}
\title{Recovering the O VII Absorption Distributions from X-Ray Data}

\author[a]{Nichole Gray}
\author[a]{Cameron T. Pratt}
\author[a]{Joel N. Bregman}
\affil[a]{University of Michigan Department of Astronomy, Ann Arbor, MI 48104, USA}

\pagestyle{empty} 
\setcounter{page}{301} 

\maketitle

\begin{abstract}
The absorption by gas toward background continuum sources informs us about the cosmic density of gas components as well as the hosts responsible for the absorption (galaxies, clusters, cosmic filaments). 
Cosmic absorption line distributions are distorted near the detection threshold (S/N $\approx 3$) due to true lines being scattered to lower S/N and false detections occurring at the same S/N. 
We simulate absorption line distributions in the presence of noise and consider two models for recovery: a parametric fitting of the noise plus a cut-off power law absorption line distribution; a non-parametric fit where the negative absorption line distribution (emission lines) is subtracted from the positive S/N absorption line distribution (flip and subtract).  We show that both approaches work equally well and can use data where S/N$\gtrsim$3 to constrain the fit.  For an input of about 100 absorption line systems, the number of systems is recovered to $\approx$14\%.
This investigation examined the O VII X-ray absorption line distribution, but the approach should be broadly applicable for statistically well-behaved data. 
\end{abstract}


\section{Introduction} \label{sec:intro}
Approximately 90\% of the baryons in the Universe remain in gaseous form \cite{fuku98,shull2012}, with about 35-50\% not yet detected with the usual absorption signatures of neutral and warm atomic gas.  This remaining gas is predicted to lie in a hotter state (T $> 5 \times 10^5$ K) that is produced by the gravitational collapse of filaments, galaxies, and galaxy clusters (e.g., \cite{sch15,hop18}).  For example, extended hot halos around galaxies are expected to have temperatures near the virial value, about $2 \times 10^6$ K for the Milky Way. Within these collapsed structures, star formation occurs, leading to the production of metals and their dispersal, typically as highly ionized outflows (e.g., \cite{BenO2018}).

Observational studies have focused on the high ionization ions of oxygen, where Lithium-like O VI  has a peak ionization fraction at $10^{5.3}$ K and is accessible in the far UV (rest wavelengths of 1032, 1038 \AA\ for the doublet).  Many UV studies have been conducted with the \textit{Hubble Space Telescope} (HST) and the \textit{Far Ultraviolet Explorer} (FUSE), showing that this ion is common in the Milky Way, halos of external galaxies, and the intergalactic medium (e.g., \cite{Wakker2003, Werk2014, Johnson2015, Qu2018Hion}).  The higher ionization ion of Helium-like O VII has a ionization fraction near unity from $0.4-2 \times 10^6$ K), while the hydrogenic form, O VIII is the dominant ion for T $> 2 \times 10^6$ K.  Exploration of these ions is only possible in the X-ray regions, as the strongest resonance lines fall at 0.574 keV (O VII) and 0.654 keV (O VIII). 

Both O VII and O VIII have been detected in the Milky Way, using the grating spectrographs on \textit{Chandra} (LETG) and \textit{XMM-Newton} (RGS) and find them to have column densities more than an order of magnitude greater than O VI \cite{Sambach2003,fang13,miller15}.  Extragalactic detection of O VII has been much more challenging, in that the equivalent widths are expected to be several times smaller than those measured for the Milky Way.  Currently, there is a singular detection, but with significant uncertainties \cite{Nicastro2018} and it will not be able to extend this effort to many sightlines, which is needed to assemble a statistically useful sample of O VII and O VIII absorption lines.  This goal can be achieved if new X-ray observatories are built with improved collecting area and spectral resolution. 

Two observatories are under development that would have the necessary capabilities – \textit{Athena} (ESA; \cite{barcons17}) and \textit{Arcus} (NASA, \cite{smith16}).  Each would detect these oxygen lines against X-ray bright AGNs at redshifts suitable to sample a useful amount of physical space so that $\sim 10^2$ absorption systems are detected \cite{breg15}.  Such studies will yield the number of absorption line systems as a function of their column density ($N$) or equivalent width ($EW$), which provide fundamental information, such as ratio of the mean ion densities to the critical density of the Universe, expressed as $\Omega$(O VII) and $\Omega$(O VIII).  There are several steps obtaining these quantities from the data, such as determining the shape of the equivalent width distribution to the smallest feasible equivalent widths.

Here we investigate the most effective approach to determining the shape of this distribution.  Toward this purpose, we adopt the observational characteristics of an \textit{Arcus}-type mission, where a spectral resolution of 3000-4500 leads to a resolution element (65-100 km s$^{-1}$; 5-7 m\AA\ at z = 0 for the O VII He$\alpha$ 21.60 \AA\ resonance line) similar to the expected width of an absorption line (45-100 km s$^{-1}$).  However, the result is rather general and can be applied to \textit{Athena} or to UV absorption line studies, provided that the spectra are well-behaved and obey the assumed Gaussian statistical characterization.

An issue that will need to be considered is contamination by Galactic absorption lines in the 10 - 40 \AA\ range, which can be estimated from theoretical models or observations toward bright Galactic X-ray sources (e.g., \cite{Miller2004,Gu2005,Gatuzz2013, Nicastro2016, Gatuzz2023}).  These lines include the standard resonance lines, plus a number of inner shell lines, which are less well modeled due to the complexity of transitions from the inner shell.  
These inner shell lines are mostly due to oxygen, an in particular O I, O II, with contributions from O III - O VI, as well as similar lines from N, Ne, and Fe.  
The inner shell oxygen lines lie longward of the O VII He$\alpha$ 21.60 \AA\ line, so they can be confused with a redshifted O VII 21.60 \AA\ feature.  There are 15 inner shell lines of consequence in the 21.60-23.55 \AA\ range, which if mistaken for redshifted O VII He$\alpha$, corresponds to a redshift range of 0 $<$ z $<$ 0.09.  The median separation between the lines is about 1.6 resolution elements for a spectral resolution of 250 (typical of a good calorimeter in this energy range, such as found in Athena), so potential IGM lines are blended with Galactic lines, making line extraction difficult if not impossible \cite{Gatuzz2023}.  A few other redshift regions would be problematic, such as at 0.30 $<$ z $<$ 0.37, and z $\approx$ 0.44. 

Soft energy grating spectrometers can have much higher spectral resolution, which mitigates this issue, such as for Arcus.  Arcus\cite{Arcus2023} is likely to have a resolution of $\approx 3500$, leading to a median of 23 resolution elements between contaminating lines.  This will lead to a modest decrease in the redshift search space in 0 $<$ z $<$ 0.09 of about 5\%.  Therefore, we will ignore this effect for the purposes of this work and concentrate on the recovery of the line distribution.

The rest of the paper is structured in the following way: \autoref{sec:motivation} gives some motivation for our work; \autoref{sec:methods} describes the methodology, specifically focusing on simulating a toy model and attempting to recover the input distribution; \autoref{sec:results} presents our results and a discussion on the implications of this work and how it can be used in future studies; and \autoref{sec:summary} summarizes the main points of this paper.

\section{Motivation} \label{sec:motivation} 

The scientific problem that motivated this study was the ability to detect the hot gas component of the Universe, using highly ionized metals, such as O VII.  The cosmological distribution of O VII absorbers as a function of EW has been calculated from numerical simulations by several authors \cite{Cen2006, Branchini2009, cen_2012, Wijers2020, Tuominen2023}. There is remarkable agreement between studies, with a similar number of absorbers per unit redshift (at low z) for N(O VII) $> 10^{15}$ cm$^{-2}$. 
Here, we used the work of \cite{cen_2012}, who calculated the distribution in the column density range of $10^{13}$ cm$^{-2}$ $<$ N(O VII) $<$ $2 \times 10^{16}$ cm$^{-2}$ (the summation that includes low temperature O VII).  The distribution is convergent at both the high and low end of the column density distribution, so to measure the total amount of N(O VII) present, one must be sensitive to sufficiently low absorption columns (we assume that the highest columns are easily measured).  This can be investigated by using the distribution of \cite{cen_2012}, where the space density per unit redshift, above some N or EW.  Here, we use EW instead of N, which for optically thin O VII is EW = 2.86 m\AA\ (N/10$^{15}$ cm$^{-2}$).  We approximated this distribution as 
\begin{equation} \label{eq:eq1}
dn(>EW)/dz = 132\, \times exp(-EW^{0.316}\,),
\end{equation}
\noindent where the EW is in units of m\AA, the units of dn($>$EW)/dz is the number of absorption systems per unit redshift with an EW above some limiting value.  

We prefer to work with the differential form of \autoref{eq:eq1}, which is the differential number of absorption systems per dEW, d$^2$n/dzdEW.
From this, one can calculate the physically meaningful total integrated EW per unit redshift in the interval of a high and low EW, dEW(EW$_{min}$, $\infty$)/dz = $\int$EW\,(d$^2n$/dzdEW)\,dEW.  This integral is evaluated from EW$_{min}$ to infinity, where EW$_{min}$ corresponds to the limiting EW for which the underlying distribution can be extracted.

\begin{figure}[t]
\centering
\includegraphics[width=0.47\textwidth]{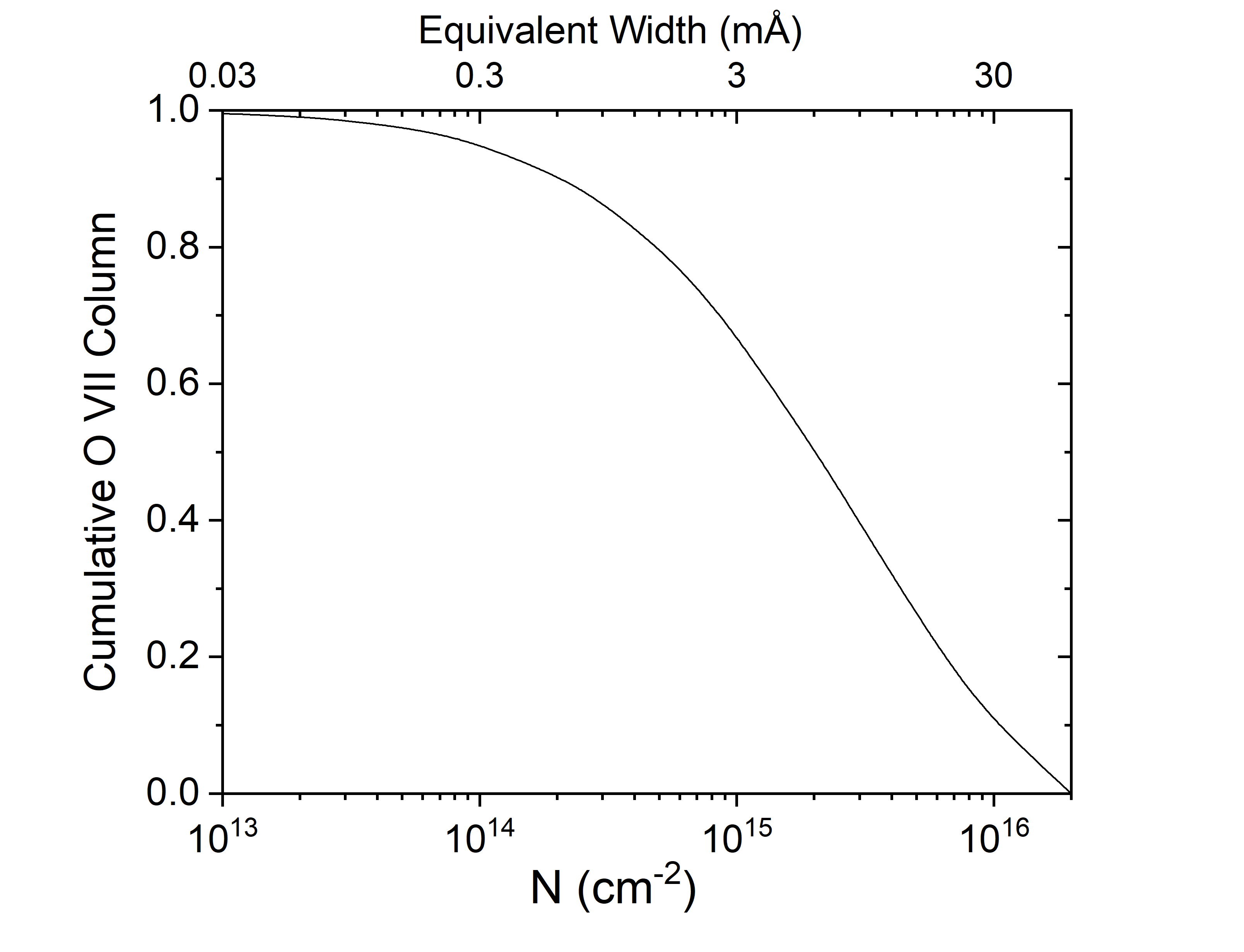}
\caption{The cumulative O VII column (or mass) as a function of the minimum limiting column of a study; the maximum column is 2$\times 10^{16}$ cm$^{-2}$ (60 m\AA).  It shows that the cumulative fractions are 50\% at 2$\times 10^{15}$  cm$^{-2}$ (6 m\AA), 75\% at 6.4$\times 10^{14}$  cm$^{-2}$ (1.9 m\AA), and 90\% at 2.1$\times 10^{14}$ cm$^{-2}$ (0.63 m\AA).
}
\label{fig:completeness}
\end{figure}

For a particular model, this quantity dEW(EW$_{min}$, $\infty$)/dz can be corrected to the value dEW(0, $\infty$)/dz, where we calculate and show this ratio,  dEW(EW$_{min}$, $\infty$)/dz\,/\,dEW(0, $\infty$)/dz in \autoref{fig:completeness}.
The quantity dEW(0, $\infty$)/dz, for O VII is proportional to the cosmologically important $\Omega$(O VII).

Our study was motivated by determining EW$_{min}$ and the ratio of dEW(EW$_{min}$,\, $\infty$)/dz\,/\,dEW(0,\, $\infty$)/dz from the observational goals of particular future X-ray missions.  
Rather than use the derivative of \autoref{eq:eq1} as the input, we use a simpler form, as the precise form varies between investigators. 

The \textit{Arcus} mission has a plan to observe approximately 30 AGN to detect approximately 100 projected absorption line systems, probing a redshift space of $\Delta z \approx 15$ \cite{smith16,Arcus2023}; \textit{Athena} has similar goals \cite{{barcons17}}. 
There will be about $30,000$ total spectral resolution elements examined in the O VII study, for the above search space, with the \textit{Arcus} resolution of 3000-4500, and with a mean uncertainty per spectral resolution element of 0.75 m\AA.  Nearly all of these resolution elements are expected to be null detections, with Gaussian noise causing each resolution to have a positive (absorption) or negative (spurious emission) value relative to the continuum. 
Thus, the challenge is to extract the underlying signal, or distribution of absorption line systems from the ensemble of values.

There are a number of issues that will arise in the actual data from a mission, such as the uncertainty being wavelength dependent, which we ignore in this work.  Also, the equivalent width of an absorption line is modified by its redshift, an issue that we do not consider, for the sake of simplicity.  These considerations can easily be dealt with in future studies. 

\section{Methods} \label{sec:methods}
\subsection{Simulated Data}

\begin{figure}[t]
\centering
\includegraphics[width=0.47\textwidth]{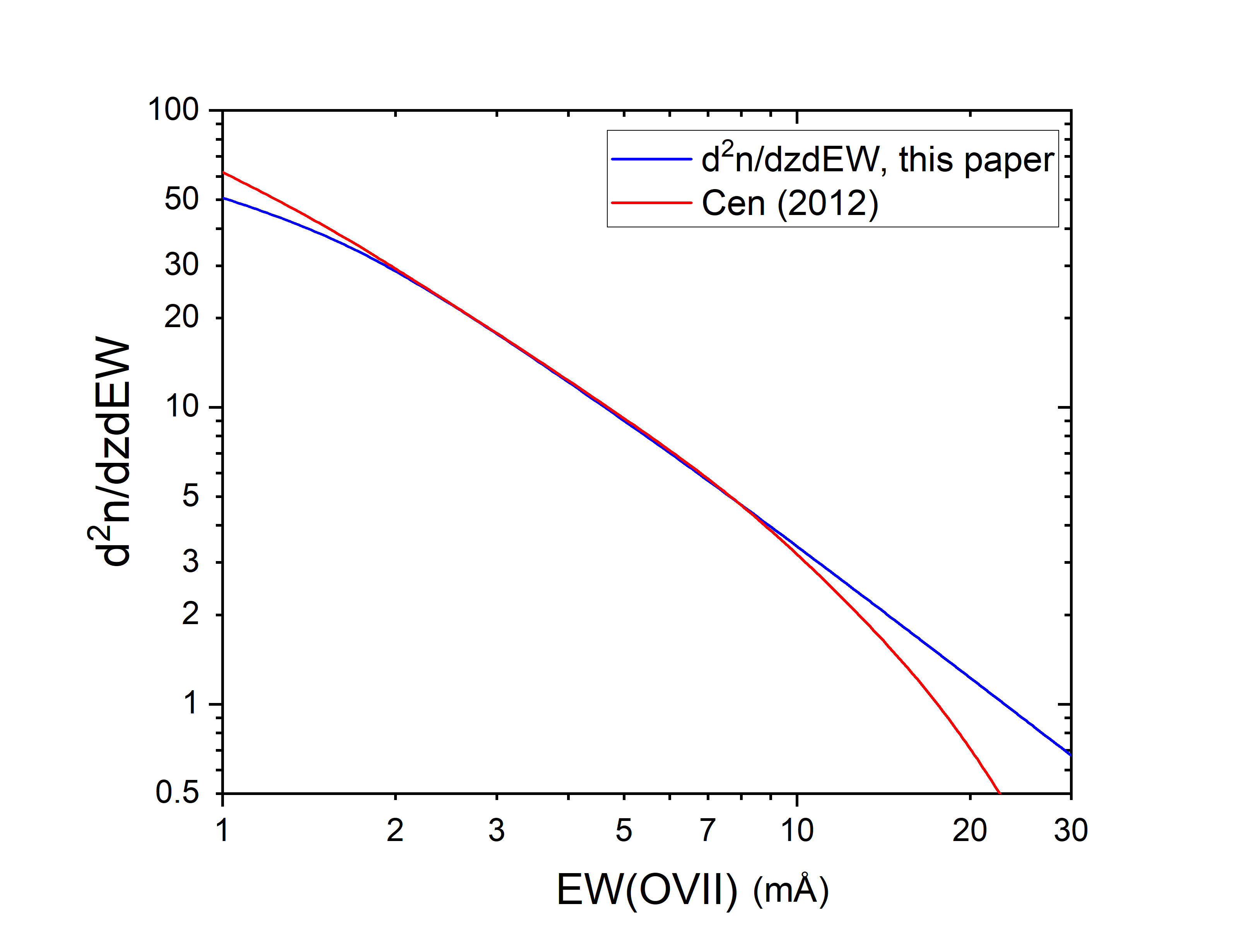}
\caption{The simplified differential column density distribution used here (eq. \autoref{eq:toy_model}, blue) compared to the distribution by \cite{cen_2012} (red), where the latter is normalized to be equal at 3 m\AA. 
The range shown here is chosen from below the EW values that provide significant constraints on the fit to the maximum EW value of the distribution. 
}
\label{fig:ToyCen}
\end{figure}

The main goal of our study is to determine the degree to which a simulated equivalent width distribution can be recovered.  We chose a distribution that captures the characteristics of the numerical simulation of \cite{cen_2012}, modified for simplicity.  
We work with the differential form of the equivalent width distribution, d$^2$n/dzdEW, where EW is the equivalent width, expressed in m\AA.  It is given by 
\begin{equation}
d^2n/dzdEW = A_{p}\left(\frac{111}{1+EW^{m}}\right),
\label{eq:toy_model}
\end{equation}

\noindent where $A_{p}$ is the normalization and $m$ is power law index. The flattening of d$^2$n/dzdEW at small equivalent width is somewhat arbitrary but does not significantly affect the determination of the two parameters, as we show below.  Also, the distribution is cut off for equivalent widths above 30 m\AA, which is motivated by the distribution of \cite{cen_2012}, as shown in 
\autoref{fig:ToyCen}. In this work, we set the normalization $A_{p} =1$ and a power-law index of $m=1.5$. 

\begin{figure}[t]
\centering
\includegraphics[width=0.47\textwidth]{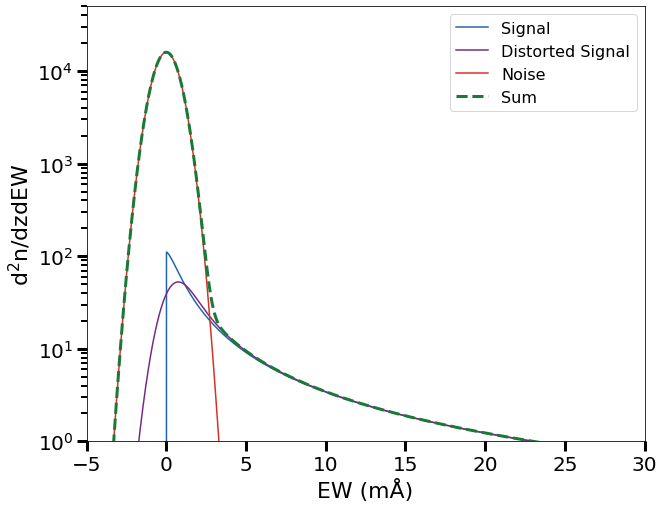}
\caption{Components of the model include the original distribution (\autoref{eq:toy_model}, blue), the distribution convolved with the Gaussian error (black), the noise from the many spectral resolution elements without absorption lines (red), and the sum (green dashed). 
}\label{fig:components}

\end{figure}

Added to this distribution is the random noise ($N_r$) in each resolution element, which forms a Gaussian ($\mathcal{N}$) centered at zero, parameterized by a normalization factor ($A_{g}$) along with a standard deviation ($\sigma_{g}$)
\begin{equation}
N_r = \mathcal{N}(A_{g},\sigma_{g}).
\label{eqn:noise}
\end{equation}
Here we assume $A_{g} = 30,000$ and $\sigma_{g} = 0.75$ m\AA, based on the baseline expectation of \textit{Arcus} \cite{smith16}.
The observed distribution, d$^2$n/dzdEW, is the convolution between the noise and true distribution, as is illustrated in \autoref{fig:components}. 
For our analysis, we performed Monte Carlo simulations, drawing $10^{3}$ random EW realizations; this number was chosen simply to be large enough in order to gather robust statistics while remaining computationally inexpensive. From each realization we obtained an EW distribution by binning the data into histograms, which we describe next.

\begin{figure*}[t!]
\centering
\includegraphics[width=.9\textwidth]{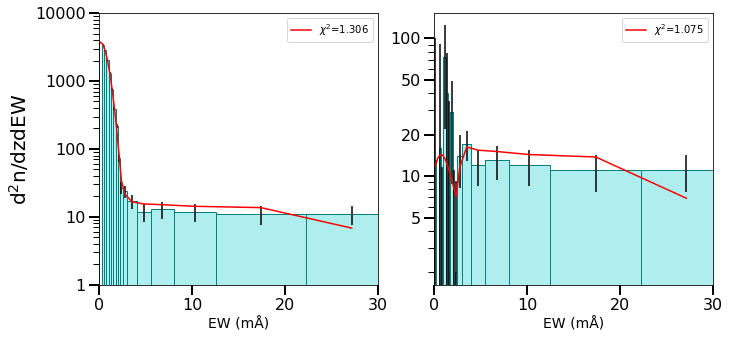}

\caption{Examples of the FPF (left) and FAS (right) recovery methods for a single realization. The best-fit model is shown in red for each and the reduced $\chi^{2}$ value is presented in the upper right of each panel. The fitting was done via least-squares regression.}\label{fig:single_realization}
\end{figure*}

\subsection{Adaptive Binning} \label{sec:bins}
Recovering simulated distributions requires binning data in order to estimate the signal properties. Unfortunately, the choice of bins is arbitrary and must be determined \textit{a priori}. In practice, one should develop a binning scheme that best represents their data i.e., one that allows for the most accurate recovery of the input signal. For example, it would not be ideal to use a linear spaced bins when trying to recover a rapidly changing distribution, which occurs here. The intrinsic EW distribution, d$^2$n/dzdEW contains few counts at large EW, so the bins are broader, while at lower EW, the bins can be smaller. It is important to use consistent binning throughout the simulations even though each realization will produce a different distribution. 

Here, we set a fiducial bin width of 0.25 m\AA\ for the low EW data. Moving toward larger EW values, we allowed the next bin to have the same width so long as the model predicted $\geq 15$ counts for that bin. This ensured that the signal-to-noise (S/N) of each bin is $\gtrsim 3$ using Poisson statistics. We acknowledge that one could use more elaborate binning schemes, such as Bayesian blocks \cite{Scargle13}, which may be considered in future work.

\subsection{Recovery Methods} \label{sec:fitmethod}

We used two methods to recover our input parameters, one being a parametric fit for the four parameters of the model, A$_p$, m, A$_g$, and $\sigma_g$ (the full-parameter fit, designated FPF).  An advantage of this approach is that we can understand the accuracy to which each parameter is fit.  A disadvantage is that we are requiring that four parameters be fit, compared to our other approach, where only two parameters are fit. 

In the other model, we did not fit for the Gaussian, but rather subtracted the absolute values of the negative EW values (apparent emission lines) from those on the positive side, our fold-and-subtract approach (FAS).  This method was non-parametric in obtaining the d$^2$n/dzdEW distribution, which was important considering it was convolved with the uncertainties. In reality, the continuum noise distribution may be more complicated than a single Gaussian defined by A$_g$ and $\sigma_g$. The FAS approach did not assume a functional form for the noise so, therefore, it would be the ideal choice for real observations. The only shortcoming to FAS is that the noise from the negative side of the Gaussian has been added to the positive side, increasing the statistical uncertainty. 

The uncertainties on the FAS result were calculated by adding the counts in the pairs of respective negative and positive bins in quadrature. Measuring counts in a bin obeys Poisson characteristics, so the uncertainty takes the form of $\nicefrac{1}{\sqrt{T}}$, with $T$ being the number of counts per bin.  Using this method, we fit for the underlying d$^2$n/dzdEW, convolved with the Gaussian error distribution. As an example, we show fits for each method given a single realization in \autoref{fig:single_realization}.

\section{Results and Discussion} \label{sec:results}

\subsection{Returned Parameter Spread} 

As mention in \autoref{sec:fitmethod}, we attempted to recover the input signal distribution using two different recovery methods. The best fit parameters for each method are presented in \autoref{tab:params}, and their distributions can be found in the corner plots in \autoref{fig:corners1} and \autoref{fig:corners2}. Both the FPF and the FAS methods recovered the input signal parameters and to similar precision. Also, both produced the same correlation between the power-law slope and normalization parameters. 

\begin{table*}
\resizebox{0.95\textwidth}{!}{%
\begin{tabular}{|ccccc|cc|}
\hline
\multicolumn{5}{|c|}{\textbf{Full Parameter Fit (FPF)}} &
  \multicolumn{2}{c|}{\textbf{Flipped and Subtract (FAS)}} \\ \hline
\multicolumn{1}{|c}{} &
  \multicolumn{1}{c|}{\textbf{A$_p$}} &
  \multicolumn{1}{c|}{\textbf{m}} &
  \multicolumn{1}{c|}{\textbf{A$_g$}} &
  \textbf{$\sigma_{g}$} &
  \multicolumn{1}{c|}{\textbf{A$_p$}} &
  \textbf{m} \\ \hline
\multicolumn{1}{|c}{} &
  \multicolumn{1}{c|}{$1.08^{+0.37}_{-0.28}$} &
  \multicolumn{1}{c|}{$1.56^{+0.15}_{-0.15}$} &
  \multicolumn{1}{c|}{$0.992^{+0.0013}_{-0.0017}$} &
  \multicolumn{1}{c|}{$0.753^{+0.0031}_{-0.0031}$} &
  \multicolumn{1}{c|}{$1.05^{+0.37}_{-0.27}$} &
  $1.56^{+0.15}_{-0.15}$ \\ \hline

\end{tabular}%
}
\vspace{0.2cm}
\caption{This table shows the average results from the full simulation and reflects the results in \autoref{fig:corners1} and \autoref{fig:corners2}. The uncertainties on these values are obtained from the $16^{th}$ and $84^{th}$ percentiles.}\label{tab:params}
\end{table*}

We note that the FPF yielded a biased estimate on the normalization of the Gaussian, $A_{g}$. Given the estimated uncertainties in \autoref{tab:params}, the value of $A_{g} = 0.992$ differs from the true value by $\approx 8\sigma$. This effect may be related to the correlations between the power-law slope and normalization parameters in \autoref{fig:corners1} and \autoref{fig:corners2}. Regardless, this bias was near the $1\%$ level and did not significantly affect our results.

\subsection{Signal-to-Noise (S/N)}\label{sec:sn}
An important empirical measurement is the S/N as a function of EW. In particular, one would like to know at what EW value should the signal separate from the noise in a significant way. In the following, we discuss how this was done using the FAS method.

The observed signal for the FAS method was obtained by subtracting the negative EW distribution from the positive side. The uncertainty for the remaining positive values were determined by simple error propagation (i.e., added in quadrature). An example of a single realization is shown in the right panel of \autoref{fig:single_realization}. One can notice the uncertainties for each bin are larger in the right panel compared to the left due to propagation of errors. 

\begin{figure}[!hb]
\centering
\includegraphics[width=0.47\textwidth]{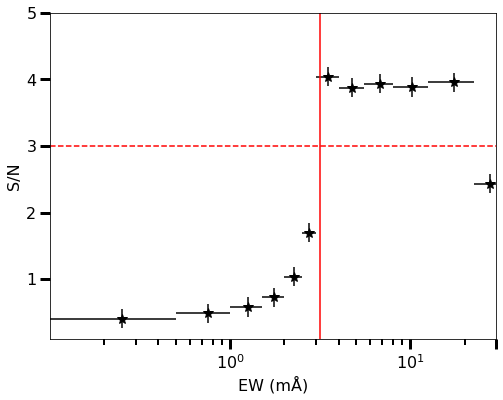}
\caption{The average S/N distribution from all realizations. The cutoff point for SN $\geq$ 3 is shown by the intersecting red lines on the left plot, with a value of $3.0 \pm 0.3$ mÅ. The horizontal error bars represent the width of each bin, and the vertical error bars denote the standard deviation using different realizations.}\label{fig:SN}
\end{figure}

The FAS method was performed over the $1000$ realizations, and the average S/N for each bin is shown in \autoref{fig:SN}. On this plot we also show a horizontal line denoting where one can achieve a S/N = 3. As expected, the S/N values were small at low EW due to the high presence of false detections. Eventually, the S/N rapidly improved and surpassed a S/N of 3 at an EW of $3.0 \pm 0.3$ m\AA\ for the FAS (similar results were obtained using FPF). In practice, this sets the theoretical limit for the minimum EW one can use to reliably measure the EW distribution in the real Universe.

Another important measured quantity was the intrinsic uncertainty in the number of detected systems\textemdash that is a cosmological variance set by the fact that we can only observe one Universe. Counting the number of detections required an estimate of the signal using either FAS or FPF. The uncertainties were large at low EW, so we measured the total number of systems above some EW threshold (here we use EW  = 3.0 m\AA). We compared the total number of systems measured above the EW threshold to the model input. The FAS and FPF methods yielded a fractional uncertainty of $\approx 14\%$. These changed to $18\%$ and $21\%$ when we lowered the EW threshold to 2 m\AA\ and 1 m\AA\ respectively. These uncertainties translate directly to the uncertainty in a cosmological density, such as $\Omega (O VII)$.

\section{Summary and Conclusion}\label{sec:summary}

The motivation for this work was to understand the uncertainties involved in determining the cosmological O VII and O VIII metal contributions to the Universe, based on X-ray missions in the planning stages, such as \textit{Athena} or \textit{Arcus}.  We adopted a functional form for d$^2$n/dzdEW that is idealized (defined by two parameters) but similar to those predicted from large scale structure simulations.  We created simulated data sets and used two approaches to recover them: a Full Parameter Fit (FPF), where we fit for both the distorted absorption line distribution, and the Gaussian noise profile; and a Flip and Subtract (FAS) method, where the distribution of negative EW features (false apparent emission features) is subtracted from the distribution of positive features (absorption lines).  In the FAS method, the net spectrum was fitted with the model d$^2$n/dzdEW that was convolved with the error distribution, which was assumed to be single Gaussian in this investigation. 

We found that both models led to equivalently good fits, and that for $\sim$100 O VII absorption systems with EW $>$ 3 m\AA\ and 30,000 null detections (but smoothed by the intrinsic Gaussian), the power-law index m was recovered to 10\% and the amplitude A$_p$ to about 30\% ($1 \sigma$).  The determination of the total EW or number of absorption systems was determined to 14\%, which sets the uncertainty in determining $\Omega$(O VII).  The fitting utilized all parts of the simulated data, but most of the power in fitting the input d$^2$n/dzdEW came from absorption features with EW $\gtrsim$ 3 m\AA. 

In an actual instrument, the sensitivity usually varies as a function of wavelength, so the noise distribution is a bit more complicated than the single Gaussian, but can still be determined by the sum of Gaussians as a function of wavelength space.  Another consideration is defining the location of the continuum, for which there is always some uncertainty.  Incorrect locations of the continuum also broaden or distort the S/N distribution, particularly at low S/N values.  An investigation of this effect is beyond the scope of this paper, but we suggest limiting the extraction of signals to $|$S/N$|$ $>$ 3 and using the FAS method.  Values of $|$S/N$|$ $<$ 3 play no role in determining A$_p$ and $m$. 

This work was meant to address a focused issue relating to particular future missions.  However, the application of this work is broader.  The approach here can be applied to determining d$^2$n/dzdEW in any wavelength region, such as for UV or optical absorption line systems, where there are already large amounts of data.

\noindent\textbf{Data Availability}

\noindent Data sharing is not applicable to this article, as no new data were created or analyzed. All data in this work was generated randomly and can be easily replicated.

\noindent\textbf{Acknowledgements}  

\noindent We would like to thank Zhijie Qu and Anne Blackwell for their comments and insights.  We gratefully acknowledge support for this work through the University of Michigan and NASA grants 80NSSC19K1013 and 80NSSC22K0481. 

\noindent\textbf{Biographies}  

\noindent Cameron Pratt is a graduate student in the Department of Astronomy at the University of Michigan.  His interests are in extragalactic astronomy, signal processing, machine learning, and statistics.

\noindent Nichole Gray is a graduate of the University of Michigan (2022) with a B.S. in Astronomy \& Astrophysics as well as in Interdisciplinary Physics. 

\noindent Joel N. Bregman is a Professor of Astronomy at the University of Michigan with interests in the gaseous content of the Universe as well as in mission planning, especially for NASA and ESA.
\clearpage

\appendix
\section{Corner Plots}
\label{appendix:A}
\begin{figure*}[h]
\centering
     \includegraphics[width=0.6\textwidth]{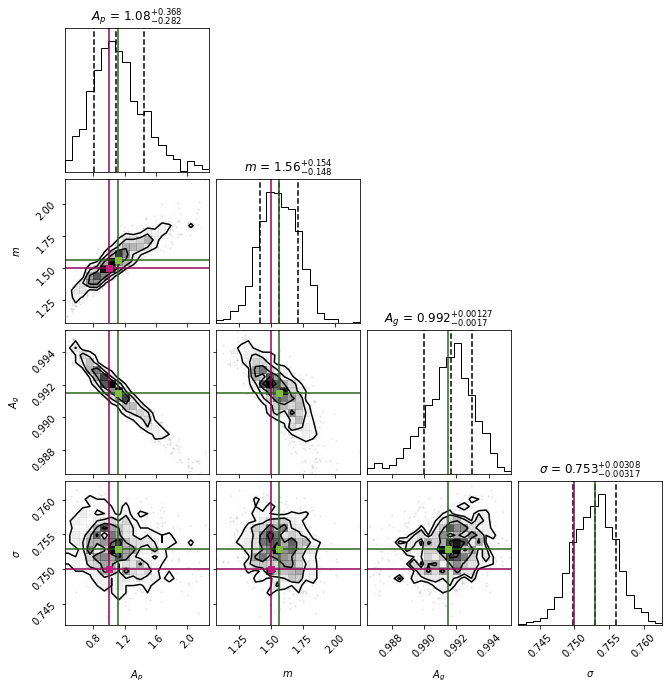}
\caption{The distribution of best-fit parameters obtained for each of the $1000$ realizations from FPF. Red lines indicate the input values and green lines give the mean. In the histograms, the dotted lines represent the $16^{th}$ and $84^{th}$ percentiles. The values above the histograms are equal to those in \autoref{tab:params}.} %
\label{fig:corners1}%
\end{figure*}

\begin{figure*}[h]
\centering

\includegraphics[width=0.4\textwidth]{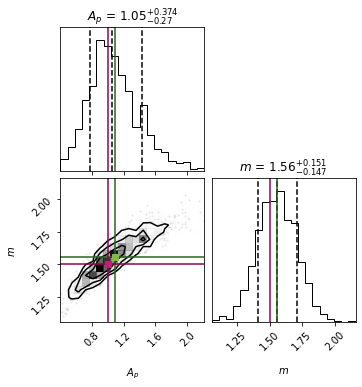}
\caption{Same as in \autoref{fig:corners1} but for the FAS method.} %
\label{fig:corners2}%
\end{figure*}

Here we present the distribution of best-fits, based on 1000 simulations in \autoref{fig:corners1} and \autoref{fig:corners2}, showing the covariance between parameters. In both FPF and FAS, there is a degeneracy between the power-law index m and the normalization of the power law, A$_p$ in the sense that a fit with a shallower power-law slope requires a larger normalization (and vice versa). 

\clearpage
\bibliography{report} 
\bibliographystyle{spiebib} 

\end{document}